\documentstyle[12pt]{article}
\input epsf

\begin{document}
\begin{titlepage}
\thispagestyle{empty}
\title{
\vspace*{-1cm}
\begin{flushright}
{\small KEK-TH-742}
\end{flushright}
\vspace{2.0cm}
On the Ambiguity of the Solution of the Muskhelishvili-Omnes Integral Equation }
\vspace{4.0cm}
\author{Tran N. Truong \\
\small \em Centre de Physique Th\'eorique, 
{\footnote {unit\'e propre 014 du
CNRS}}\\ 
\small \em Ecole Polytechnique \\
\small \em F91128 Palaiseau, France \\
 and \\
\small \em Institute of Nuclear and Particle Studies, \\
\small \em High Energy Accelerator Research Organization,\\
\small \em 1-1 Oho, Tsukuba, Ibaraki, 305 - 081 Japan } 

\date{January 2001}

\maketitle

\begin{abstract}

The solution of the Muskhelishvili-Omnes Integral Equation is ambiguous by a real polynomial. The
coefficients of this polynomial can be fixed  either by the knowledge of the low energy parameters
or by the asymptotic behavior of the form factor. The role of the contact terms of
the low energy effective Lagrangian is  explicitly analysed.

\end{abstract}
\end{titlepage}

There have been discussions of contact terms in the phenomenological models describing low energy
pion phenomena \cite{rudaz, schecter, bando}.
They seem to play an important role in analysing the pion form factor
and the process $\gamma \pi^0 \rightarrow \pi^+\pi^-$. The main difference between this model and
the usual vector meson model is the existence of a contact term where the photon interacts
directly with the charged pions and not by the intermediate of the vector $\rho$ meson.
Phenomenological analysis of this contact term in the form factor has recently been given
\cite{benayoun}.

 The purpose of
this note is to analyze the existence of the contact term within the spirit of the more general
approach to the form factor problem using the  Muskhelishvili-Omnes (MO)
integral equation \cite{omnes}. It is based on the assumption of the elastic unitarity and
dispersion relation for the pion form factor. The main conclusion of our analysis is that the
existence of the contact term is in conflict with unitarity in the form of the phase theorem. Its
existence can, however, be simulated by a polynomial ambiguity  contribution to the 
 solution of the MO equation. This polynomial is interpreted physically as either the correction due
to the inelastic contribution to the unitarity relation or the modification of the
asymptotic behavior of the form factor.

Let us first discuss the solution of this problem from a more general viewpoint. The form factor
$V(s)$ is an analytic function with a cut from $4m_\pi^2$ to $\infty$.

Assuming that $V(s)$ is polynomially bounded and that $V(s)s^{-(n+1)} \rightarrow 0$ as $s
\rightarrow \infty$, $n>0$, we  can write the following subtracted dispersion relation:
\begin{equation}
V(s)=1+a_1s+...a_{n}s^{n}+ \frac{s^{n+1}}{\pi}\int_{4m_\pi^2}^\infty
\frac{ImV(z)dz}{(z)^{n+1}(z-s-i\epsilon)}
\label{eq:dr} 
\end{equation}
where we have used the Ward identity $V(0)=1$.
Around $s=0$, the dispersion integral is of the order
$(s)^{n+1}$ and can be neglected, hence the low energy theorem is recovered. Needless to
say, $a_n$ are, apart from a factorial factor $n!$, the derivatives of $V(s)$ evaluated at
$s=0$. These could be taken from experiments or could be determined by sum rules using some
assumptions on the asymptotic behavior of the form factor. 

For example, if we assume that as $s\rightarrow\infty$, $V(s)/s \to 0$, a once subtraction dispersion
relation for $V(s)$ can be written; the first and higher derivatives of $V(s)$  can be written as:
\begin{equation}
V^{(n)}(0) = \frac{n!}{\pi}\int_{4m_\pi^2}^\infty \frac{ImV(z)dz}{z^{n+1}} \label{eq:der}
\end{equation}

The mathematical problem is now  clear: Find the solution of the integral
equation of the MO type for $V(s)$ with  its imaginary part given by the elastic
unitarity:
\begin{equation}
ImV(s)= V(s) e^{-i\delta (s)} \sin \delta (s) \label{eq:imo}
\end{equation}
 with the boundary conditions around $s=0$ given by Eq. (\ref{eq:dr}) and where $\delta$ is the
the strong P-wave $I=1$ $\pi\pi$ phase shifts.

To solve this integral equation Eq. (\ref{eq:dr}) which is of MO type
\cite{omnes},  let us define the function  
$\Omega(s)$  normalized to unity at $s=0$:
\begin{equation}
\Omega(s)= \exp(\frac{s}{\pi} \int_{4m_\pi^2}^\infty
\frac{\delta(z)dz}{z(z-s-i\epsilon)})
\label{eq:omn}
\end{equation}
 The solution for our integral equation is:
\begin{equation}
V(s) = P_n(s)\Omega(s) \label{eq:solo}
\end{equation}
where $P_n(s)$ is a  polynomial in $s$ of order n with real coefficients, $P_n(0)=1$. 
From eq. (\ref{eq:solo}), the phase of the form factor $V(s)$ is $\delta(s)$ which is also  a
direct consequence of the unitarity relation \cite{watson}. It will be called in the following as
the phase theorem. 

 The ambiguity of the solution of
the MO equation due to the polyomial is not surprising because its solution  depends on the knowledge
of the asymptotic behavior of the form factor. 
 The polynomial could be interpreted as
extra parameters which are introduced in the problem without violating the phase theorem; in many
cases they can be determined  using experimental data at low energy.

This is indeed true by expanding the function
$\Omega(s)$ in a power series in
$s$,
  and compare it with the derivatives given in Eq. (\ref{eq:solo}) and
Eq. (\ref{eq:dr}). The expansion of $\Omega(s)$ in the Taylor's series around $s=0$ is possible
because  
$\Omega(s)$  is an analytic function with a cut from $4m_\pi^2$ to $\infty$.

In the special case where only two terms in the series are known, i.e. the pion charge by Ward's
indentity and its first derivative $a_1$ which is proportional to the r.m.s. pion radius by
experiment, the solution of our integral equation is given  by:

\begin{equation}
V(s)= \{1+ s(a_1-\Omega^{'}(0)) \}\Omega(s) \label{eq:sol}
\end{equation}
where $\Omega^{'}(0)$ denotes the first derivative of $\Omega(s)$ evaluated at $s=0$
and is given by:
\begin{equation}
\Omega{'}(0)=\frac{1}{\pi} \int_{4m_\pi^2}^\infty \frac{\delta(z)dz}{z^2}
\label{eq:dero}
\end{equation}
The presence of the term $\Omega^{'}(0)$ is to ensure the boundary condition for
$V(s)$ is  satisfied. If $a_1-\Omega^{'}(0)=0$ there is no need to introduce a momnomial term in
the form factor and hence the asymptotic behavior of the form factor is solely determined by
$\Omega(s)$. (Using partial fraction for the dispersion integral together with the sum rules Eq.
(\ref{eq:der}) the asymptotic behavior of $V(s)$ can be determined).

 It is straightforward to generalise the solution of Eq. (\ref{eq:dr}) for other values of
$n$. For example when  $n=2$, the solution for the integral equation Eq.( \ref{eq:dr}) is
obtained by adding to the curly bracket on the righthand side of Eq. (\ref{eq:sol}) a 
term:
\begin{equation}
s^2(
a_2-a_1\Omega^{'}(0)-a_0\frac{\Omega^{''}(0)}{2}+\Omega^{'2}(0))
\label{eq:n=2}
\end{equation}

Let us now look at the experimental data. For a given set of P-wave $\pi\pi$ phase shifts $\delta$,
one can, in principle, construct the function $\Omega(s)$ as given by Eq. (\ref{eq:omn}) by the 
numerical method. We prefer instead to construct this function  using  the inverse amplitude
method or the [0,1] Pad{\'e} approximant method for the one loop Chiral Perturbation Theory for the
pion form factor and verify that the resulting phases are in agreement with
experiments \cite{truong1}. The function
$\Omega(s)$ can be parametrised as follows:
\begin{equation}
         \Omega(s) = \frac{1} {1 -s/s_{R} - {1\over
96\pi^2f_\pi^2}\{(s-4m_\pi^2)
 H_{\pi\pi}({s}) + {2s/3}\}} \label{eq:vu1}
\end{equation}
where $f_\pi=0.093 GeV$, and $s_{R}$ is related to the $\rho$ mass squared $m_\rho^2$ by 
requiring that the real part of the denominator of Eq. (\ref{eq:vu1}) vanishes at the $\rho$
mass; $H_{\pi\pi}({s})$ is a well-known integral over the phase space factor:
\begin{equation}
   H_{\pi\pi}(s) = (2 - 2\sqrt{1-4m_\pi^2/ s}) \log{\frac{\sqrt{s}+\sqrt{s-4m_\pi^2}}{
2m_\pi}}+i\pi\sqrt{1-4m_\pi^2/s} \label{eq:H}
\end{equation}
for $ s>4m_\pi^2$; for other values of s, $H_{\pi\pi} (s)$ can be obtained by analytic
 continuation. $\Omega(s)$ as given by Eq. (\ref{eq:vu1}) has a ghost at a very large negative value
of $s$ with a small residue and is irrelevant for our low energy calculation.

 Although
we do not try the best fit to the experimental data on the P-wave
$\pi\pi$ phase shifts,  an adequate agreement is obtained between theory and experiment as shown in
Fig. (1) where $m_\rho=0.775 GeV$.

Eq. (\ref{eq:vu1}) can also be obtained by the vector meson dominace model \cite{gounaris} with the
strong
$\rho\pi\pi$ coupling constant given by the KSRF relation \cite{ksrf}.

If it was assumed that there is no polynomial ambiguity 
$P_n(s)=1$, we would have:
\begin{equation}
V(s)=\Omega(s) \label{eq:na}
\end{equation}

Fitting this expression with the physical value of the $\rho$ mass, the pion rms radius calculated
by this expression is too low by more than
$10\%$ and the maximum value of the form factor squared at the $\rho$ peak (the $\rho$ leptonic
width) is too low by
$30\%$, Fig.(2). This implies that the assumed elastic unitarity relation is not good enough. One can
either rewrite the integral equation taking into account of the inelastic effect which was
previously studied \cite{truong2} or to make more subtractions in the integral equation and keeping
the elastic unitarity relation, in order to minimize the high energy contribution. We choose first
the latter method. (The two-loop inverse amplitude or the [0,2] Pad{\'e} approximant method are
given recently and are found to give excellent agreement with the pion form factor data
\cite{hannah}).

 Let us write a twice subtracted dispersion relation, including the inputs as the pion
charge and the experimental value of the pion radius. The integral equation now reads:
\begin{equation}
V(s) =1+ a_1 s
+\frac{s^2}{\pi}\int_{4m_\pi^2}^\infty
\frac{V(z)e^{-i\delta(z)}\sin\delta(z)dz}{z^2(z-s-i\epsilon)}
\label{eq:rms}
\end{equation}
where $a_1=1/6(<r^2>)$ with $<r^2>$ is the rms squared radius of the pion. The factor $z^2$ in the
denominator of the dispersion integral suppresses the high energy contribution. 

 Using the experimental
value
$<r^2>=0.439\pm.008 fm^2 $ \cite{na7} , the solution of the integral solution is given by Eq.
(\ref{eq:sol})
 and is numerically equal to \cite{truong3, truong4}
\begin{equation}
V(s)= (1+0.15s/m_\rho^2)\Omega(s) \label{eq:num}
\end{equation}
The $\rho$ leptonic width and the square of the modulus of  the pion form factor are now in agreement
with the experimental data, Fig.(2).

The other possibility to get agreement with experimental data  is to introduce the inelastic
effect which could occur at higher energy. The integral equation now reads:
\begin{equation}
V(s)=1+ \frac{s}{\pi}\int_{4m_\pi^2}^\infty dz
\frac{f^*(z)V(z)+\sigma(z)}{z(z-s-i\epsilon)} \label{eq:ine}
\end{equation}
where $f(s)=(2i)^{-1}(\eta e^{2i\delta(s)}-1)$, $\eta$ is the inelastic factor ($\eta=1$ for
elastic $\pi\pi$ scattering) and $\sigma$ is the inelastic spectral function due to the
contribution of higher intermediate states in the unitarity relation. The solution of this integral
equation is known:
\begin{equation}
V(s)=\Omega(s)\{1+\frac{s}{\pi}\int_{s_i}^\infty dz
\frac{2Re(\sigma(z)e^{i\delta(z)})}{(1+\eta(z))e^{-i\delta(z)}\Omega(z)}\frac{1}{z(z-s-i\epsilon)}\}
\label{eq:moin}
\end{equation}
$s_i$ is the inelastic threshold and for the pion form factor problem one can take it to be
around $1-1.3 GeV^2$. In reality, it could be as high as 2 $GeV^2$ in the $\rho{'}$ resonance where
the inelastic effect would become important.

The dispersion integral in Eq. (\ref{eq:moin}) is seen to be an analytic
function with a cut starting at $s_i$ to $\infty$ and hence it can be expressed as a Taylor's
series around $s=0$. The radius of convergence of this series is $s_i$. Hence  the RHS of Eq.
(\ref{eq:moin}) is simply the solution of the MO equation as given by Eq. (\ref{eq:solo}) and the
fit to the experimental data is given by Eq. (\ref{eq:num}). 

For a feel for the inelastic effect, let us see how does the $\rho{'}(1500)$ ,  which dominantly decays
into
$4\pi$, influence the pion form factor. The inelastic spectral function can then be approximated by a
delta function in the dispersion integral which yields a term proportional to $\Omega(s)$
multiplying with the
$\rho{'}$ propagator. This assures the validity of the phase theorem below the inelastic threshold
and that neither the $\rho{'}$ propagator alone nor a pure contact term can be written. For $s$ in
the $\rho$ region, one can expand the $\rho^{'}$ propagator in a series in $s$ and fit to the
experimental data to get the desired result Eq. (\ref{eq:num}).

Let us now compare our result with models with a contact term \cite{rudaz, schecter} and, in
particular, the hidden local symmetry model \cite{bando}. At the tree level, the hidden local
symmetry model gives:
\begin{equation}
V^{HLS}(s) =-\frac{a}{2}+1+\frac{a}{2}\frac{m_\rho^2}{m_\rho^2-s} \label{eq:hs1}
\end{equation}
which can be rewritten as:
\begin{equation}
V^{HLS}(s)	     =\frac{m_\rho^2+s(a/2-1)}{m_\rho^2-s} 
 \label{eq:hs1bis} 
\end{equation}
In order to make the $\rho$ unstable we could try to replace  the factor $m_\rho^2/(m_\rho^2-s)$
in Eq. (\ref{eq:hs1}) by $\Omega(s)$  leaving the contact term unchanged. This would violate the
phase theorem unless the constant term cancels each other to give a net result which is equal to
$\Omega(s)$; this solution does not fit with the experimental data as explained above. 

We  should use instead Eq. (\ref{eq:hs1bis}) and interprete it as  the limit large
$s$,
$V^{HLS}\rightarrow (1-a/2)$ (in the limit of zero width).  Then the hidden symmetry solution for
the vector form factor must contain a polynomial
$P_1(s)$ or:
\begin{equation}
V^{HLS}=(1+\frac{a/2-1}{m_\rho^2}s)\Omega(s) \label{eq:hs2}
\end{equation}
Compare this equation with Eq. (\ref{eq:num}) we have $a=2.30$. Within the validity of the above
interpretation of the contact term, this result is in agreement with that obtained by the authors of
the reference \cite{benayoun} who begin with Eq. (\ref{eq:hs1}), then make the $\rho$ unstable. The
unitarisation scheme is done with the $N/D$ method for the P-wave $\pi\pi$ scattering.

The limit of large value of $s$ of $V^{HLS}$ should be taken with some reservation, because  the
inelastic effect interpretation of the polynomial ambiguity is only valid for $s<s_i$,  Eq.
(\ref{eq:moin}) and $s_i$ is experimentally not very large.  

Another example of the contact term \cite{rudaz,schecter, bando} is discussed in the $\gamma
\pi^0 \to \pi^+\pi^-$ process, in 
particular, the hidden symmetry model \cite{bando}. In the VMD model, the
$\gamma\pi^0
\rightarrow
\pi^+\pi^-$ is written as:
\begin{equation}
A^{3\pi}(s,t,u) =\frac{\lambda}{3}(\frac{m_\rho^2}{m_\rho^2-s}+\frac{m_\rho^2}{m_\rho^2-t}+
\frac{m_\rho^2}{m_\rho^2-u}) \label{eq:3p1}
\end{equation}
where $s,t,u$ are the standard invariant kinematics. With a contact term it  can be written as:
\begin{equation}
A^{3\pi}(s,t,u)
=\frac{\lambda}{3-c}\{\frac{m_\rho^2}{m_\rho^2-s}+\frac{m_\rho^2}{m_\rho^2-t}+
\frac{m_\rho^2}{m_\rho^2-u}-c\} \label{eq:3p2}
\end{equation}
where $c$ is proportional to the strength of the contact term and $\lambda$ is given by the anomaly:
\begin{equation}
\lambda=\frac{e}{4\pi^2 f_\pi^3}=  9.7  GeV^{-3} \label{eq:anomaly3}
\end{equation}
Eq. (\ref{eq:3p2}) can be rearranged to give:
\begin{equation}
A^{3\pi}(s,t,u)
=\frac{\lambda}{3}\{[\frac{m_\rho^2}{m_\rho^2-s}(1+\frac{c}{3-c}\frac{s}{m_\rho^2})]+[s\to t] +[s
\to u]\}
\label{eq:3p3}
\end{equation}

Eq. (\ref{eq:3p1}) yields a decay width $\Gamma(\rho\to \pi\gamma)=36 KeV$ which is too small
compared with the experimental data. Eq. (\ref{eq:3p2}) with  $c=1$  yield, on the other hand, a
width which is 9/4 times larger and appears to be in better agreement with the data \cite{caparo,
huston}.

To make the $\rho$ unstable,  we could make a
substitution
$m_\rho^2/(m_\rho^2-s)$ with
$\Omega(s)$  in Eq. (\ref{eq:3p1}) and Eq. (\ref{eq:3p2}), then both resulting equations would
violate the phase theorem. Following the spirit of the form factor calculation, one could try to
make this replacement in Eq. (\ref{eq:3p3}) but the resulting equation still violates the phase
theorem when the P-wave amplitude is projected out from the full amplitude. How to get a correct
result which satisfies the elastic unitarity relation or the phase theorem
\cite{watson}, is a lenghthy problem and will be dealt in a forthcoming publication \cite{truong5}.
We would like to point out however the ambiguity of the corresponding integral equation, which is a
generalisation of the MO equation, also exists but its nature is more complicated than that given in
Eq. (\ref{eq:sol}). 

We have shown in this note two methods to get agreement with experimental data. The first one
consists in making a twice subtractions in the MO integral equation in order to suppress the high
energy contribution in the dispersion integral and hence it allow us to use the approximation of the
elastic unitarity for the pion form factor. The second method is to take into account of the
inelastic effect which, around the $\rho$ region, can be simulated as the polynomial ambiguity of
the solution of the MO equation. In both methods, the extra parameter is determined by the (low
energy) rms radius of the pion and hence we can only calculate the $\rho$ leptonic width and
also its total width Eq. (\ref{eq:vu1}). (It should be noticed that the rms pion radius can be
calculated by a rapidly converging sum rule in terms of the magnitude of the pion form factor factor
and the P-wave $\pi\pi$ phase shifts using Eq. (\ref{eq:der}) with $n=1$ \cite{truong4}). 

Within the frame work of dispersion relation and unitarity which gives rise to the MO equation,
there is no room for a constant contact term. Its existence could, however, be introduced as a large
energy behavior of the form factor or some higher energy inelastic effect as explained in this
note. 

This work is completed at the Institute for Nuclear and Particle Studies at the KEK High Energy
Accelerator Research Organization. The author would like to thank Professors M. Kobayashi
and H. Sugawara for hospitality.

\newpage

{\bf Figure Captions}

Fig.1~: The phases of the function $\Omega(s)$ or $V(s)$ in degrees are given as a function of
$s(GeV^2)$. The experimental data are taken from references \cite{proto, hyams, martin}.

Fig.2~:The square of the modulus of the function $\Omega(s)$ is given as a function of $s$ in
$GeV^2$ (dashed line). The square of the modulus of the pion form factor $V(s)$ as given by Eq.
(\ref{eq:num}) (solid line). Experimental data are taken from references \cite{barkov, aleph}.

\newpage
\begin{figure}
\epsfbox{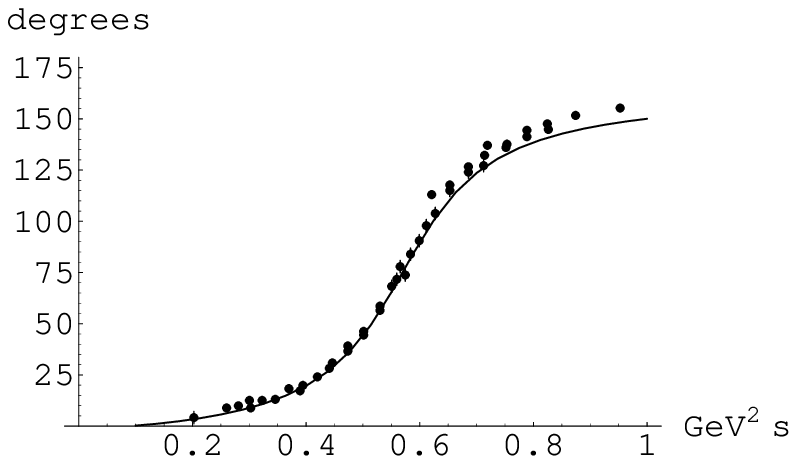}
\caption{}
\label{Fig.1}
\end{figure}

\begin{figure}
\epsfbox{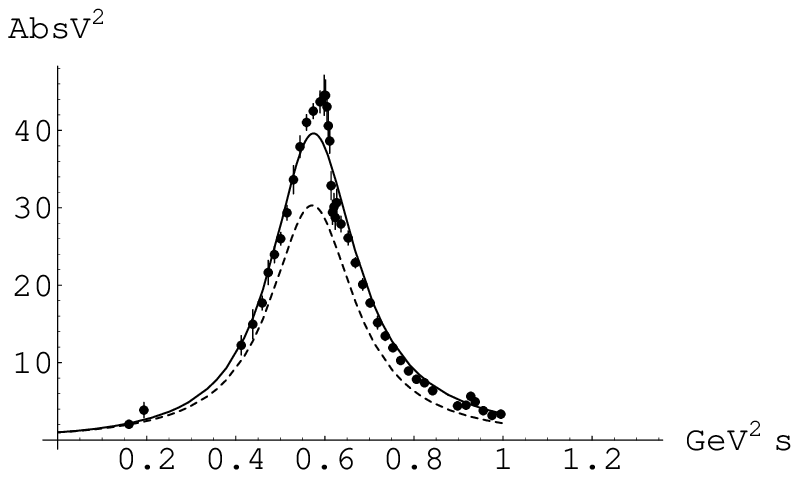}
\caption{}
\label{Fig.2}

\end{figure}

\end{document}